\documentclass[a4paper,11pt]{article}

\pdfoutput=1 % if your are submitting a pdflatex (i.e. if you have
\usepackage{soul}
\usepackage{jcappub} % for details on the use of the package, please
\newcommand{\orcidlink}[1]{\href{https://orcid.org/#1}{\textcolor{gray}{\includegraphics[height=1.7ex]{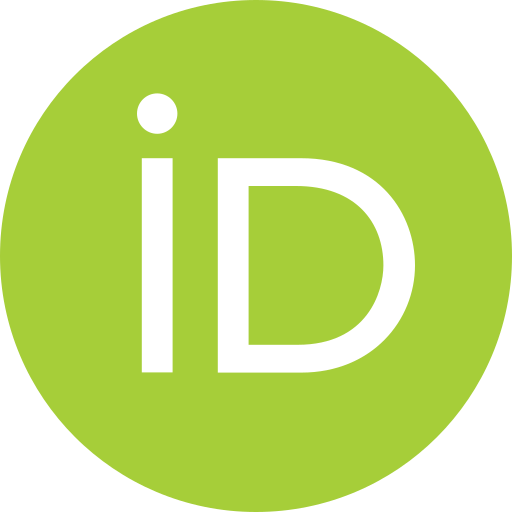}}}}

\author{Sourav Roy Chowdhury$^{a}$\orcidlink{0000-0003-2802-4138};Aritra Basak$^{b}$; Maxim Khlopov$ ^{c}$\orcidlink{0000-0002-1653-6964}; Maxim Krasnov$ ^{d, e}$\orcidlink{0009-0003-1740-5397}}
\affiliation{		%$ ^1 $ Research Institute of Physics, Southern Federal University, 344090 Rostov on Don, Russia.\\
		$ ^a $ Department of Physics, Vidyasagar College, 39, Sankar Ghosh Lane, Kolkata, India.\\
            $ ^b $  Department of Physics, Ramakrishna Mission Vivekananda Educational And Research Institute, G. T. Road, Po Belur Math, Howrah, West Bengal 711202.\\
		$ ^c $ Virtual Institute of Astroparticle Physics, 75018 Paris, France.\\
		$ ^d $ National Research Nuclear University “MEPHI,” 31 Kashirskoe Chaussee, 115409 Moscow, Russia.\\
$ ^e $ Research Institute of Physics, Southern Federal University, 344090 Rostov on Don, Russia
}
\emailAdd{souravrc79@gmail.com}   
\emailAdd{aritrabasak12@gmail.com}
\emailAdd{khlopov@apc.in2p3.fr}   
\emailAdd{morrowindman1@mail.ru}

\usepackage[T1]{fontenc} % if needed
\usepackage{lineno}

\title{\boldmath Inferring the Merger History of Primordial Black Holes from Gravitational-Wave data and the Stochastic  Signatures.}
\abstract{
    Primordial black holes (PBHs) are well-motivated candidates for cold dark matter and may also account for a fraction of the binary black hole mergers observed by the LIGO-Virgo-KAGRA Collaboration. In this study, we investigate the gravitational-wave signatures of PBHs, with a particular focus on evaluating their integrated contribution to the stochastic gravitational-wave background arising from binary mergers over a broad range of redshifts. We perform a Bayesian analysis of gravitational-wave events following all Gravitational-Wave Transient Catalog data, assuming a log-normal PBH mass function. We compute the merger rate distribution of PBH binaries by accounting for gravitational torques from the surrounding PBH. To constrain this rate, we employ the latest limits from the third observing run of LIGO/Virgo. Owing to their primordial origin, PBHs exhibit enhanced merger activity at high redshifts, prior to the onset of stellar formation. Our analysis yields a relatively weak inference on the redshift evolution index of the PBH merger rate, with $\alpha = 2.19^{+0.16}_{-0.16}$ at 68\% confidence level. The local merger rate of PBH binaries is found with posterior estimates lying in the range $23.5-30.3~\mathrm{Gpc}^{-3}\,\mathrm{yr}^{-1}$, reflecting a high degree of statistical precision in the inferred distribution. Additionally, we emphasize the potential of stochastic gravitational-wave background observations to probe the cumulative history of PBH mergers across cosmic time.
}

\date{}
\usepackage{xcolor}

\begin{document}

\maketitle

\section{Introduction}

The detections of gravitational waves (GWs) from binary black hole (BBH) mergers, black hole-neutron star mergers and binary neutron star (BNS) mergers by the LIGO-Virgo-KAGRA (LVK) observatories \cite{Abbott_2016,bhns,bns1,bns2} have started the era of GW and multimessenger astronomy. Numerous BBH merger events have since been observed \cite{LIGOEvents1,LIGOEvents2,LIGOEvents4,Abbott_2021,Abbott_2023,Gwtc3} ; however, the origin of their progenitor systems remains an open question. A variety of formation channels—ranging from isolated binary evolution \cite{mt2,mt1}, dynamical interactions in dense stellar environments \cite{dyn_evo2,dyn_evo3}, chemically homogeneous evolution \cite{che_evo2,che_evo3} and even primordial black hole (PBH) scenarios—have been proposed to explain the observed BBH population \cite{PBH_fc1,PBH_fc2}. Current measurements constrain the local BBH merger rate to 17.9–44 Gpc$^{-3}$ yr$^{-1}$ \cite{Gwtc3}.

Cosmological sources, expected to be associated with inflationary dynamics, provide a unique window into the physics of the primordial Universe, whereas astrophysical contributions are instrumental in probing high-redshift source populations and assessing the existence of PBH binaries. In addition to individually resolvable GW events, the stochastic gravitational-wave background (SGWB) arises from the superposition of both resolved and unresolved sources \cite{sgwb1,sgwb_det}. Background is termed ``stochastic'' due to its intrinsically statistical nature. Although several astrophysical and cosmological phenomena—such as core-collapse supernovae, magnetars, cosmic strings, relic GW from inflation—may contribute to SGWB \cite{ccsne1,ccsne2,magnetars1,magnetars2,string1,string2,infla1,infla2}, it is expected to be dominated by compact binary coalescences, particularly BBH mergers. SGWB is characterised by the dimensionless energy density spectrum $\Omega_{\mathrm{GW}}(\nu)$, which is typically estimated through cross-correlation analyses of strain data from multiple detectors \cite{sgwb1,sgwb_met}. Results from the first three observing runs (O1–O3) of the LVK Collaboration yielded no statistically significant detection, but placed an upper limit of $\Omega_{\mathrm{GW}}(25~\mathrm{Hz}) \leq 1.04 \times 10^{-9}$ \cite{Gwtc3} under the assumption of a power-law spectrum with spectral index $2/3$, consistent with predictions from BBH-dominated SGWB models. In contrast, recent results from several pulsar timing array collaborations- PPTA \cite{ppta}, NANOGrav \cite{nano}, EPTA/InPTA \cite{epta}, and CPTA \cite{cpta}—have reported evidence for an isotropic stochastic signal.

The study of black holes (BHs), celestial objects characterised by gravitational fields so intense that no matter or radiation can escape, represents a cornerstone of modern astrophysics and cosmology. While astrophysical BHs, formed through the gravitational collapse of massive stars, have been extensively investigated, there exists an intriguing theoretical alternative: PBHs \cite{Bird_2016, Sasaki:2016jop, Chen_2018,Liu_2019,Chen_2022}. PBHs are hypothesised to have originated in the early universe, offering a unique probe of extreme physical conditions shortly after the Big Bang. The concept of PBHs was initially explored in the seminal works of Zel'dovich and Novikov \cite{PBH1} and later on by others \cite{Hawking, carr}. These investigations posited that in the primordial universe's extreme density and inhomogeneity, significant density fluctuations could have led to localised regions collapsing under their self-gravity, directly forming BHs. This formation mechanism contrasts fundamentally with the stellar evolutionary processes responsible for astrophysical BH genesis. Beyond the well-studied mechanisms involving matter field overdensities, there is also a possibility that PBHs could be formed without explicit involvement of matter fields. These alternative pathways are significant as they highlight the potential role of primordial geometry and quantum gravitational effects in the very early universe. As demonstrated in \cite{10.3389/fspas.2022.927144}, PBHs could arise purely from the dynamics of compact extra space in quadratic $f(R)$-gravity with tensor corrections. This mechanism exploits instabilities inherent in the modified gravitational theory and the compactification process, resulting in the direct gravitational collapse of spacetime regions into black holes. Such scenario offer distinct signatures and constraints compared to matter-driven formation. Comprehensive reviews encompassing both standard and these alternative PBH formation mechanisms can be found in Refs. \cite{particles6020033, Sasaki_2018,Green_2021}.

While PBHs remained largely speculative for decades, recent theoretical developments and observational capabilities have revitalised their status as viable dark matter candidates. Various scenarios have been proposed in which PBHs originate from the collapse of large density perturbations generated in the early universe, with models predicting their formation across a wide range of cosmological epochs and resulting in diverse mass spectra that reflect the underlying early-universe physics~\cite{Maxim,Garc_a_Bellido_2017,Belotsky,Franciolini_2022, cur-pbh3}. However, recent advancements in both theoretical modelling and observational techniques have established robust frameworks for explaining their formation~\cite{Sasaki_2018, Liu_2019b, Cai_2019, Cai_2020,Jedamzik_2020,Carr_2021, Carr_2020, Liu_2023, Wang_2023}. In particular, the potential existence of PBHs within the stellar-mass window has gained considerable attention, as their mergers may constitute a non-negligible contribution to both current and future GW detections~\cite{De_Luca_2021, Vaskonen_2021, De_Luca_2021b, H_tsi_2021, cur_pbh, cur-pbh1, cur-pbh2}. Abundance of PBHs in cold dark matter (CDM), denoted as \( f_{\text{PBH}} \), is estimated to be on the order of \( \mathcal{O}(10^{-3}-10^{-1}) \) to explain the BBH events observed by the LVK collaboration~\cite{Liu_2023, Escrivà_2023}. Accurately estimating the merger rate distribution of PBH binaries is essential for extracting PBH population parameters from GW data, especially given that factors such as hierarchical mergers and the binary formation history can significantly influence the observed mass distribution~\cite{Jangra_2023,sgwb5m}. 
%Notably, the merger rates predicted by PBH models are consistent with the rates inferred from GW observations, further supporting the hypothesis that these events may involve PBH binaries.

Among emerging tools to distinguish PBH from (Astrophysical BH)ABH, the SGWB offers unique sensitivity to early-universe phenomena. Joint PBH-ABH population models~\cite{sgwb1m,sgwb2m} demonstrate that the SGWB carries information about merger histories and redshift evolution. In clustered PBH environments, gravitational perturbations suppress wide binary mergers, inducing a spectral turnover from the canonical \( \Omega_{\text{gw}} \propto \nu^{2/3} \) to \( \Omega_{\text{gw}} \propto \nu^{-65/28} \)~\cite{sgwb5m}. Current and forecasted SGWB measurements further constrain PBH abundance over a broad mass range~\cite{sgwb3m}. Additionally, realistic modelling of PBH binaries within halos and subhalos reveals that substructure and clustering significantly enhance the SGWB amplitude and alter its spectral shape, with subhalo contributions dominating the high-frequency regime \cite{sgwb4m}. Measuring eccentricity in BBH mergers offers a promising observational method to distinguish PBH origins from conventional astrophysical channels in GW detections \cite{eccn}.

% current work focuses on PBHs and their GW signatures, with a particular emphasis on the SGWB; where the PBH binary merger rate is significantly shaped by third-body interactions. This study aims to model the redshift-dependent merger rate density of PBHs, constrain their population parameters using observational data, and evaluate their cumulative contribution to the SGWB. 

Our current work focuses on PBHs and their GW signatures, with particular emphasis on quantifying the cumulative contribution of PBH mergers to the SGWB across a range of redshifts. A central feature of our framework is the inclusion of third-body interactions, which induce angular momentum in close PBH pairs, thereby facilitating binary formation and significantly influencing merger rates and timescales.
Nonetheless, these studies do not account for the scenario in which a PBH binary merges into a resultant BH/PBH, which subsequently forms a new binary system through interaction with another PBH. We adopt a log-normal mass distribution for the PBH population, characterized by the central mass \( M_c \), width \( \sigma \), abundance fraction \( f_{\mathrm{PBH}} \), and redshift evolution index \( \alpha \). The merger rate density is modelled as a function of redshift, and the detector-frame rate is computed by integrating over the comoving volume while accounting for cosmological effects. To infer the underlying PBH population parameters, we employ a hierarchical Bayesian inference framework using BBH merger data from the GWTC catalog. The likelihood is constructed from the predicted redshift-dependent merger rate, and posterior distributions are obtained using the \texttt{Bilby} framework with a nested sampling algorithm. Observational priors are imposed to ensure consistency with local merger rate constraints. In the second part of the study, we compute the SGWB spectrum \( \Omega_{\mathrm{GW}}(f) \) arising from unresolved PBH binary mergers by integrating the energy spectra over cosmic history. The predicted background is evaluated against the sensitivity curves of current and future GW detectors, including Advanced LIGO, Virgo and the Einstein Telescope.

Our paper is organised as follows: In Section \ref{sec2}, we present the theoretical framework for modelling the merger rate density of PBH binaries. In Section~\ref{sec3}, we estimate the model parameters through Bayesian inference using data from the O1–O3 observing runs, with particular emphasis on the estimation of the PBH mass distribution and abundance. In Section \ref{sec4}, we compute the SGWB from unresolved PBH mergers and discuss its observational prospects. We finally present our results in Section \ref{sec5} and conclude in Section \ref{sec6}.

\section{Model of the Merger Rate Density Distribution of PBH Binaries} \label{sec2}

A rigorous analysis of the merger rate density of PBH binaries requires a comprehensive understanding of the gravitational mechanisms that drive their coalescence throughout cosmic history. The merger history effect plays a central role in determining the dynamical evolution of PBH binaries and carries significant implications of GW observations. Such analysis is grounded in the theoretical framework describing the PBH mass distribution and its contribution to the CDM density, providing essential context for assessing the cosmological impact of PBH mergers.

The mass distribution of PBHs is represented by a mass function, denoted as $P(m)$, which is normalized by the following expression:

\begin{equation}
    \int_0^\infty P(m) \, dm = 1
\end{equation}
ensuring total probability is conserved.

When evaluating the contribution of PBHs to CDM, the abundance of PBHs within a mass interval \( (m, m + \mathrm{d}m) \) is given by~\cite{Chen_2018,Chen_2019}:
\begin{equation}
0.85\, f_{\text{PBH}}\, P(m)\, \mathrm{d}m,
\end{equation}
where \( f_{\text{PBH}} \) denotes the fraction of CDM composed of PBHs, and the numerical factor 0.85 accounts for the non-relativistic matter component, including both CDM and baryons.

Following \cite{Liu_2019,Liu_2019b}, the effective PBH mass, \( m_{\text{PBH}} \), is defined as:

\begin{equation}
\frac{1}{m_{\text{PBH}}} = \int \frac{P(m)}{m} \, \mathrm{d}m,
\end{equation}
which characterises the statistical expectation over the mass distribution. 

The fraction of PBHs with mass $m$ in the present-day average number density, relative to the total average PBH number density, is defined in Ref. \cite{Liu_2019} as

\begin{equation}
F(m) = \frac{P(m)}{m_{\text{PBH}}} m,
\end{equation}
which preserves the normalisation of the distribution across the PBH population.

To compute the merger rate density of PBH binaries, it is assumed that PBHs are randomly distributed after matter-radiation equality, following a spatial Poisson process~\cite{Nakamura_1997,Sasaki:2016jop}. Gravitational interactions between nearby PBHs lead to the formation of binaries that eventually merge via GW emission. The merger rate density from the first-merger process, \(\mathcal{R}(t, m_i, m_j)\), is then obtained by integrating over the mass $m_l$, of the third PBH that induces the binary's angular momentum~\cite{Liu_2019b,Liu_2023b}.

The merger rate density for PBH binaries is computed by integrating over the mass \( m_l \) of the third PBH, which induces angular momentum in the binary system:

\begin{equation}
\mathcal{R}(t, m_i, m_j) = \int R_l(t, m_i, m_j, m_l)\, \mathrm{d}m_l,
\end{equation}
where the integrand \( R_l \) is given by~\cite{Liu_2019b, Liu_2023b}

\begin{equation}
R_l(t, m_i, m_j, m_l) = 1.32 \times 10^6 \left(\frac{t}{t_0}\right)^{-\frac{34}{37}} \left(\frac{f_{\text{PBH}}}{m_{\text{PBH}}}\right)^{\frac{53}{37}} m_l^{-\frac{21}{37}} (m_i m_j)^{\frac{3}{37}} (m_i + m_j)^{\frac{36}{37}} F(m_i) F(m_j) F(m_l),
\end{equation}
with \( t \) denoting cosmic time and \( t_0 \) the current age of the universe.

The total merger rate density is obtained by integrating over all possible binary masses:

\begin{equation}
\mathcal{R}(t) = \int \mathcal{R}(t, m_i, m_j)\, \mathrm{d}m_i\, \mathrm{d}m_j.
\end{equation}

This theoretical framework enables the prediction of the GW background arising from PBH mergers, thereby providing a range of PBH population parameters. It supports the interpretation of some observed BBH events by LVK as potentially primordial in origin. Hierarchical Bayesian inference techniques further refine these estimates by incorporating observational data and PBH mass functions into a statistically robust analysis.

Cosmic time \( t \) and redshift \( z \) are treated as interchangeable variables, and their interrelation is specified by

\begin{equation}
t(z) = \int_z^{\infty} \frac{dz'}{(1+z') H(z')},
\end{equation}
where $H(z)$ is the Hubble parameter as a function of redshift.

\section{Bayesian Inference of PBH Population Parameters} \label{sec3}

To estimate the merger rate density of PBH binaries, it is essential to accurately model both the population parameters and the underlying mass distribution of PBHs. A widely adopted model for the PBH mass distribution is the log-normal form, which is particularly suitable for describing PBHs in the inflationary power spectrum. The corresponding probability density function (PDF) for the PBH mass \( m \) is given by:

\begin{equation}
P(m) = \frac{1}{\sqrt{2 \pi} \, \sigma \, m} \exp \left[-\frac{(\ln(m / M_c))^2}{2 \sigma^2} \right],
\end{equation}
where \( M_c \) denotes the characteristic mass at the peak of the distribution, and \( \sigma \) quantifies the spread. This normalized distribution plays a central role in capturing the mass variability of PBHs formed in the early universe.

We perform Bayesian parameter estimation to constrain the properties of the PBH population using GW merger data. We consider the normalized likelihood function evaluated over the theoretical PBH merger rate density $\mathcal{R}_{\mathrm{PBH}}(m_1, m_2, z)$, parameterized by the PBH mass function, the abundance fraction $f_{\mathrm{PBH}}$, and the redshift evolution index $\alpha$. This framework allows us to evaluate how well a particular set of model parameters can describe the observed BBH mergers identified by the LVK collaboration.

For a set of population parameters \( \Lambda = \{ M_c, \sigma, f_{\mathrm{PBH}}, \alpha \} \), we evaluate the likelihood of $N$ observed BBH merger events with measured component masses and redshifts, for the dataset \( d = \{ m_1^{(i)}, m_2^{(i)}, z^{(i)} \}_{i=1}^{N_{\mathrm{obs}}} \) as:
\begin{equation}
    \mathcal{L}(d \mid \Lambda) \propto \prod_{i=1}^{N_{\mathrm{obs}}} \frac{\mathcal{R}_{\mathrm{PBH}}(m_1^{(i)}, m_2^{(i)}, z^{(i)} \mid \Lambda)}{\mathcal{N}(\Lambda)},
\end{equation}
where \( \mathcal{R}_{\mathrm{PBH}}(m_1, m_2, z \mid \Lambda) \) is the redshift-dependent PBH merger rate density predicted by the model and \( m_1^{(i)} \) and \( m_2^{(i)} \) denote the primary and secondary black hole masses (with \( m_1^{(i)} \geq m_2^{(i)} \)), and \( z^{(i)} \) is the redshift of the \( i \)-th event. \( \mathcal{N}(\Lambda) \) is a normalization factor that ensures the likelihood integrates consistently over the accessible parameter space:
\begin{equation}
    \mathcal{N}(\Lambda) = \int_{m_1^{\mathrm{min}}}^{m_1^{\mathrm{max}}} \int_{m_2^{\mathrm{min}}}^{m_2^{\mathrm{max}}} \int_0^{z_{\mathrm{max}}} \mathcal{R}_{\mathrm{PBH}}(m_1, m_2, z \mid \Lambda) \, \Theta(m_1 \geq m_2) \, dz \, dm_2 \, dm_1,
\end{equation}
where $\Theta(m_1 \geq m_2)$ enforces the physical ordering of component masses. The priors implemented for the system are provided in the Table \ref{tab:PBH_param_summary}.

To maintain consistency with observational constraints, we restrict the parameter space to combinations that produce a local merger rate within the empirically established range of $17.9$–$44$~Gpc$^{-3}$~yr$^{-1}$ at redshift $z = 0.2$ \cite{Gwtc3}, thereby ensuring the physical viability of the model. The inference framework also rejects combinations that produce non-positive rates or result in invalid normalization.

\begin{table}[h]
\centering
\renewcommand{\arraystretch}{1.3}
\begin{tabular}{ c l l l}
\hline
\textbf{Parameter} & \textbf{Description} & \textbf{Prior Distribution} & \textbf{Posterior} \\
\hline
$M_c$ & Central mass scale & $\mathcal{U}(10,~ 50)$& $21.44^{+0.79}_{-0.77}$ \\
$\sigma$ & Width of the distribution & $\mathcal{U}(0,~ 1)$ & $0.84^{+0.03}_{-0.03}$ \\
$\log_{10} f_{\mathrm{PBH}}$ & PBH abundance (log scale) & $\log_{10} \mathcal{U}(-4, ~-2)$ & $-2.67^{+0.01}_{-0.01}$ \\
$\alpha$ & Redshift evolution index & $\mathcal{U}(1,~ 3)$ & $2.19^{+0.16}_{-0.16}$ \\
\hline
\end{tabular}
\caption{Model parameters for Log-normal PBH mass function, including their physical interpretations, prior distributions, and posterior estimates.}
\label{tab:PBH_param_summary}
\end{table}

\begin{figure}
    \centering
    \includegraphics[width=0.75\linewidth]{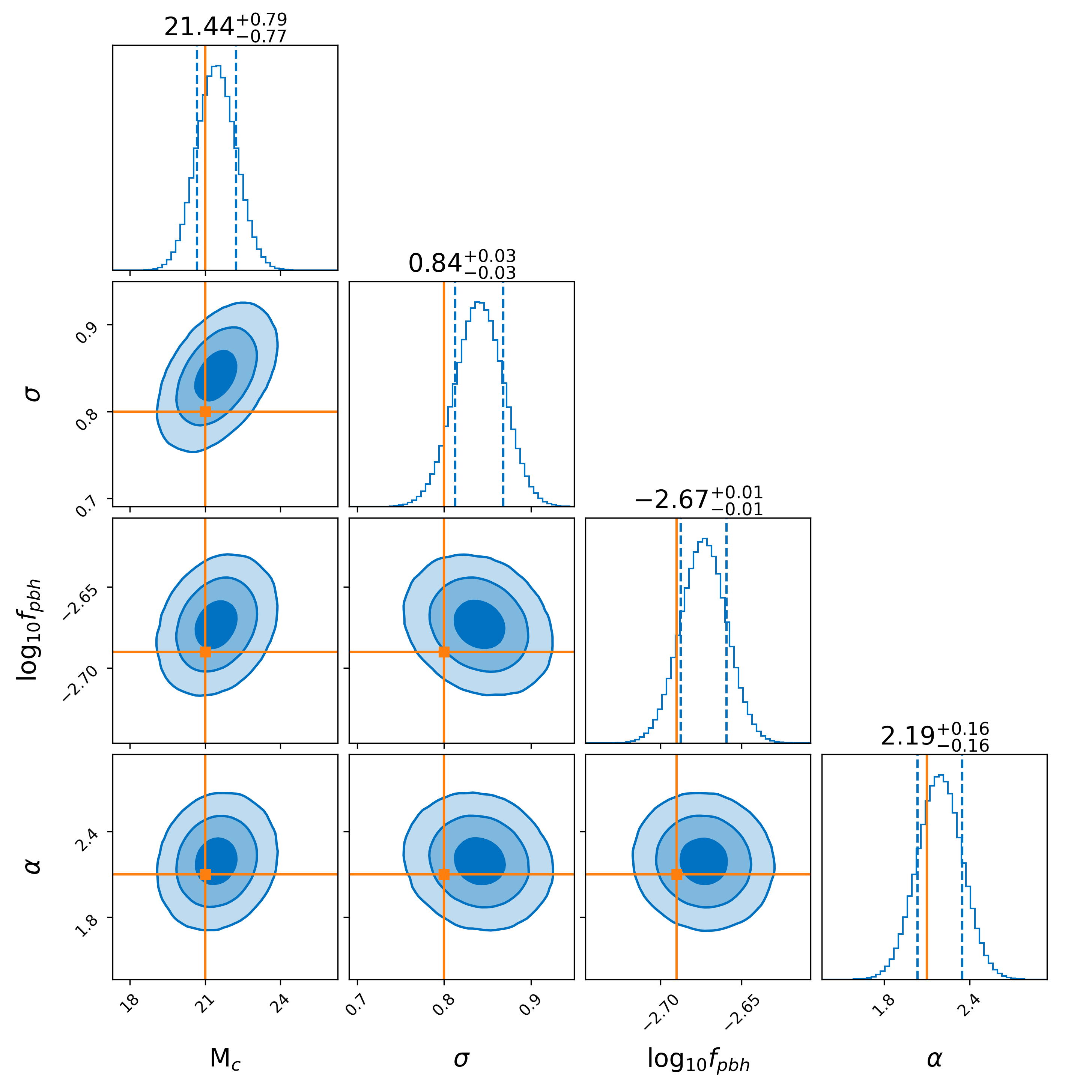}
    \caption{The marginalized one- and two-dimensional posterior distributions for the parameters ${M_c, \sigma, f_{\rm PBH}} {\rm~and~} \alpha $, assuming a log-normal mass function, as inferred from the GWTC catalogs. The posterior was obtained using a nested sampling algorithm implemented in the \texttt{Bilby} framework. Orange vertical and horizontal lines indicate the injected parameter values.}
    \label{f1}
\end{figure}

Bayesian inference allows us to quantify the degree of belief in different sets of model parameters \( \Lambda = \{ M_c, \sigma, f_{\mathrm{PBH}}, \alpha \} \), given the data \( d \), by computing the posterior probability distribution:
\begin{equation}
    p(\Lambda \mid d) \propto \mathcal{L}(d \mid \Lambda) \, \pi(\Lambda),
\end{equation}
where $\pi(\Lambda) = \pi(M_c) \cdot \pi(\sigma) \cdot \pi\left(\log_{10} f_{\mathrm{PBH}}\right) \cdot \pi(\alpha)$ denotes the joint prior distribution. The final posterior distributions for $M_c$, $\sigma$, $f_{\mathrm{PBH}}$, and $\alpha$ characterize the PBH mass function and its contribution to the observed BBH merger rate, as shown in Fig.~\ref{f1}.

The posterior distribution is computed using the \texttt{Bilby} \cite{bilby} framework with the \texttt{dynesty} nested sampler. This approach provides a computationally efficient and statistically robust method for comparing PBH merger models against observational data, without requiring the full machinery of hierarchical reweighting or detailed modelling of selection effects.

\begin{figure}
    \centering
    \includegraphics[scale=0.8]{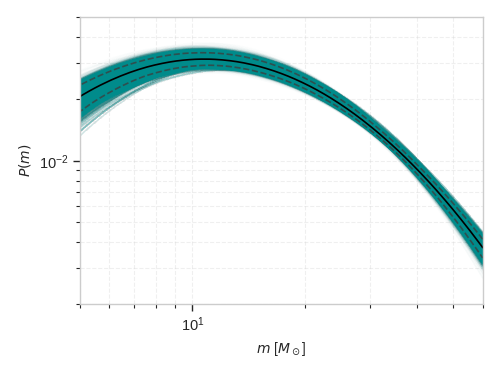}
    \caption{Posterior predictive distributions for PBHs assuming a log-normal mass function, inferred from the GW Transient catalog via Bayesian analysis. The black solid line denotes the median, and dashed lines indicate the 90\% credible interval.}
    \label{pm}
\end{figure}

Figure \ref{pm} presents the posterior predictive distributions (PPDs) for the log-normal PBH mass functions. The PPD quantifies the probability of parameter values \( \theta \) given the observed data \( d \), synthesizing prior knowledge with empirical evidence derived from the data. The hyperposterior \( p(\Lambda \mid d) \) represents the posterior probability distribution for the hyperparameters \( \Lambda \) that characterise the shape of the mass distribution \( P(\theta) \), thus encapsulating our updated statistical understanding of the PBH population model.

\section{Stochastic Background from Compact Binary Coalescences} \label{sec4}

SGWB arising from the coalescence of compact binary systems is generated by the superposition of numerous sources that are individually undetectable by a given network of GW detectors. The combined signal from these unresolved sources gives rise to a diffuse background, whose energy density—when normalized by the critical energy density of the Universe, \( \rho_c c^2 \)—defines the dimensionless spectral energy density parameter \( \Omega_{\rm GW}(f) \)~\cite{omega1}:

\begin{equation}
\Omega_{\text{GW}}(f) = \frac{1}{\rho_c} \frac{d\rho_{\text{GW}}}{d\ln f},
\end{equation}
where \( \rho_{\text{GW}} \) is the energy density of GWs per logarithmic frequency interval and \( \rho_c = \frac{3 H_0^2}{8\pi G} \) is the critical energy density of the universe, with \( H_0 \) being the Hubble constant. This quantity represents the fractional energy density in GW relative to the critical density.

For a stochastic background originating from a population of BBH mergers, the spectral shape of \( \Omega_{\text{GW}}(f) \) can be computed by integrating the redshift-dependent merger rate and the GW energy spectrum emitted by each event \cite{sgwb1,sgwb_det}. The expression reads
\begin{equation}
\Omega_{\text{GW}}(f) = \frac{f}{\rho_c H_0} \int_0^{z_{\max}} \frac{\mathcal{R}_{PBH}(z)}{(1+z) E(\Omega, z)} \frac{dE_{\text{GW}}}{df_s} dz,
\end{equation}
where \( \frac{dE_{\text{GW}}}{df_s} \) is the GW energy spectrum emitted by a binary system in the source frame with \( f_s = (1+z)f \), and the factor in the denominator arises from the cosmological volume element. The function $E(\Omega, z)$ is defined as,
$ E(\Omega, z) = \sqrt{\Omega_r (1+z)^4 +\Omega_m (1+z)^3+\Omega_\Lambda}, $
where, $\Omega_r = 9.094 \times 10^{-5}$, $\Omega_m = 0.315 \pm 0.007$, $\Omega_\Lambda = 0.685 \pm 0.007$ and $H_0 = 67.66 \pm 0.42$ km/s/Mpc are the standard $\Lambda$CDM parameters \cite{planck}, represent the density parameters for radiation, pressureless matter, cosmological constant and the Hubble constant, respectively.

\begin{figure}
    \centering
    \includegraphics[scale=0.8]{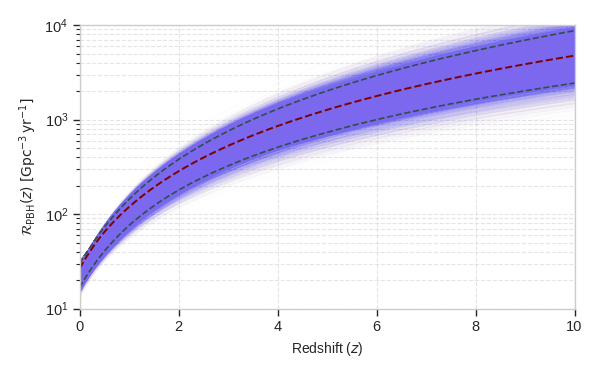}
    \caption{Redshift evolution of the PBH merger rate density, $\mathcal{R}_{\rm PBH}$(z), inferred using a log-normal mass function from GWTC data. Blue curves show posterior samples, the maroon dashed line denotes the median, and black dashed lines indicate the 90\% credible interval.}
    \label{rpbh}
\end{figure}

The redshift evolution of the PBH merger rate is expected to exhibit an increasing trend with redshift, reflecting the higher merger probability at earlier cosmic epochs. To parameterise this behaviour, we adopt a phenomenological power-law form \cite{sgwb1m}:
\begin{equation}
\mathcal{R}_{\rm PBH}(z) = \mathcal{R}_{\rm PBH}(0, m_1, m_2)(1 + z)^\alpha,
\end{equation}
where \( \mathcal{R}_{\mathrm{PBH}}(0, m_1, m_2) \) denotes the local (i.e., \( z = 0 \)) merger rate for binaries with component masses \( m_1 \) and \( m_2 \), and \( \alpha > 0 \) is the redshift evolution index. The parameter \( \alpha \) is treated as a free variable, enabling a flexible and model-independent characterisation of PBH contributions to the SGWB. The PPDs of the merger rate for binaries is represented in Figure \ref{rpbh}.

The energy emitted per unit frequency in the source frame, \( \frac{dE_{\mathrm{GW}}}{df_s}\), is expressed in terms of the source properties and encapsulates the distinct frequency dependence during the inspiral, merger, and ringdown phases, following the phenomenological waveform model of Ajith et al.~\cite{Ajith} as: 

\begin{equation}
\frac{dE_{\mathrm{GW}}}{df_s} = A\,
\begin{cases}
 f_s^{-1/3}, & f_s < f_{\mathrm{merge}} \\[2ex]
 f_{\mathrm{merge}}^{-1} f_s^{2/3}, & f_{\mathrm{merge}} \leq f_s < f_{\mathrm{ring}} \\[3ex]
 \dfrac{f_s^2 f_{\omega}^4}{f_{\mathrm{merge}} f_{\mathrm{ring}}^{4/3} \left[4(f_s - f_{\mathrm{ring}})^2 + f_{\omega}^2\right]^2}, & f_{\mathrm{ring}} \leq f_s < f_{\mathrm{cut}}
\end{cases}
\end{equation}
where $A = \frac{(\pi G)^{2/3} M^{5/3} \eta}{3}$, total mass of a binary \( M = m_1 + m_2 \) and symmetric mass ratio \( \eta = m_1 m_2 / M^2 \). The characteristic transition frequencies are the merger frequency \( f_{\mathrm{merg}} \), ringdown frequency \( f_{\mathrm{ring}} \), and cutoff frequency \( f_{\mathrm{cut}} \)---are given by
\begin{equation}
f_x = \frac{c^3 (a_1 \eta^2 + a_2 \eta + a_3)}{\pi G M}, \label{eq45}
\end{equation}
where \( f_x \) denotes any of \( f_{\mathrm{merg}} \), \( f_{\mathrm{ring}} \), or \( f_{\mathrm{cut}} \); \( G \) is Newton's gravitational constant; and \( a_1, a_2, a_3 \) are phenomenological coefficients specified in Table~1 of Ajith et al.~\cite{Ajith}. The Lorentzian width parameter for the ringdown, \( f_{\omega} \), is defined similarly as in Eqn. (\ref{eq45}).

\begin{figure}
    \centering
    \includegraphics[scale=0.85]{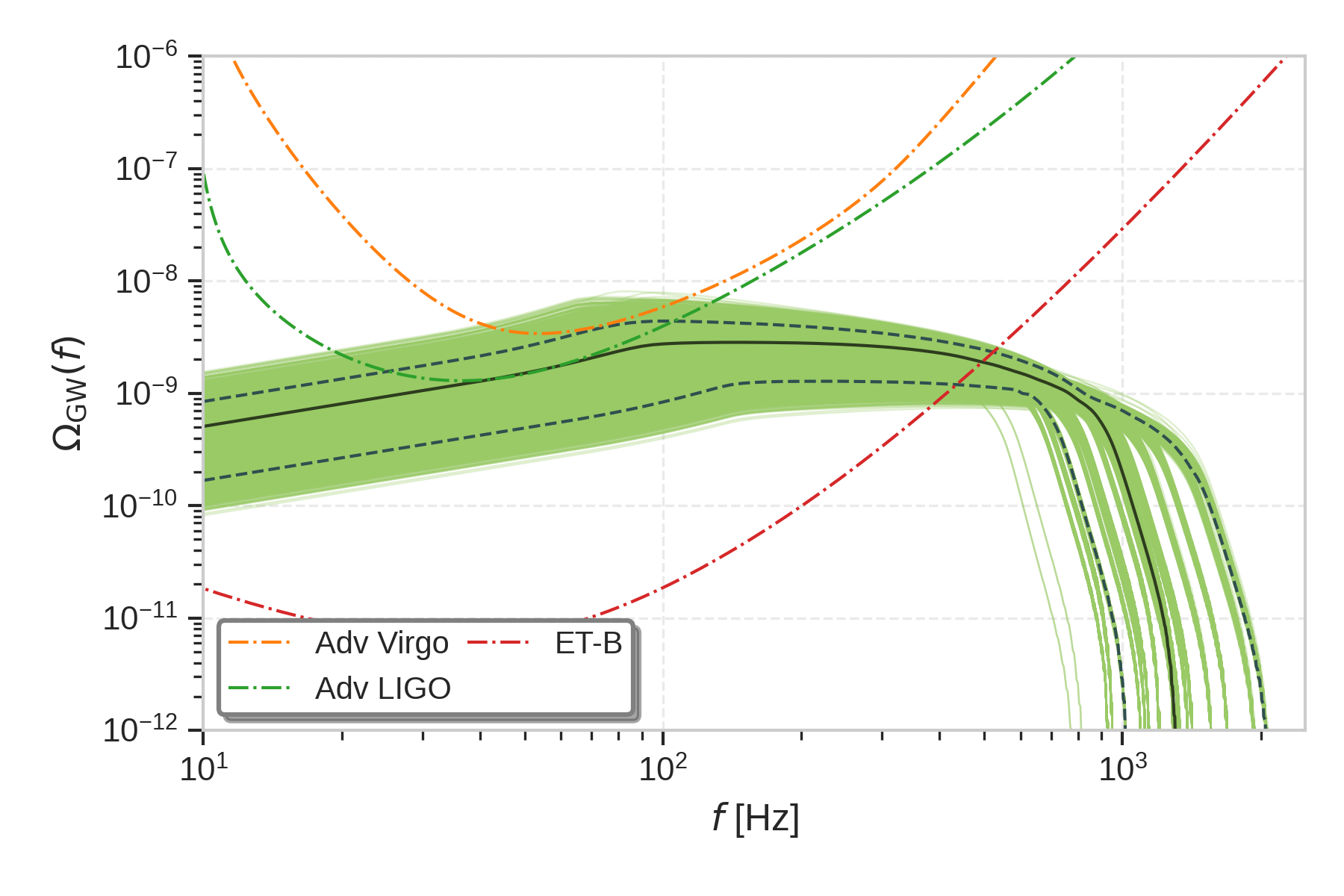}
    \caption{Dimensionless energy density spectra of the SGWB, $\Omega_{\mathrm{GW}}(f)$, generated by a population of BBH mergers. The solid blue line represents the median background spectrum, while the dashed brown lines indicate the 5 and 95th percentile intervals. The sensitivity curves of current and future GW detectors—Advanced LIGO, Advanced Virgo and ET, with SNR = 2 and a year of exposure —are overlaid for reference. }
    \label{sgwb}
\end{figure}

The frameworks establish a connection between early-universe processes responsible for the emission of GWs and the signals accessible to current and future detectors. In the context of PBH scenarios, both the frequency spectrum and amplitude of \( \Omega_{\text{GW}}(f) \) are sensitive to the underlying PBH mass distribution, the redshift evolution of merger rates, and the specific channels of binary formation. As such, measurements of the SGWB provide a complementary means of probing the potential contribution of PBHs to the dark matter content of the Universe. The variation of the energy density spectra \( \Omega_{\text{GW}}(f) \) is shown in Figure \ref{sgwb}. The dashed curves correspond to the SGWB computed for networks including Advanced Virgo, LIGO, and the Einstein Telescope (ET-B), respectively. The SGWB spectra and SNR forecasts assume one year of observation, an SNR threshold of 2. For the purposes of this analysis, the overlap reduction function \( \gamma(f) \) is set to 1 for co-located detector pairs (such as Advanced LIGO and Virgo), and to \(-3/8\) for the ET-B configuration, following Refs .~\cite {omega2,omega3}.

Intrinsic detector noise can mask or mimic a SGWB, making detection challenging. Signal-to-noise ratio (SNR) provides a quantitative measure of the background’s strength relative to the noise, and is a critical benchmark for detectability \cite{omega1}. Detection of SGWB relies on cross-correlating the outputs of pairs of detectors. SNR for such a measurement, assuming Gaussian stationary noise, is given by
\begin{equation}
\text{SNR}^2 = 2T \int_0^\infty df \, \frac{\gamma^2(f) \Omega_{\text{GW}}^2(f)}{f^6 P_1(f) P_2(f)} \left( \frac{3 H_0^2}{10 \pi^2} \right)^2,
\end{equation}
where \( T \) is the observation time, \( P_1(f) \) and \( P_2(f) \) are the one-sided noise power spectral densities of the two detectors, and \( \gamma(f) \) is the overlap reduction function, which encodes the loss in sensitivity due to separation and relative orientation.

\section{Results}\label{sec5}

Based on a log-normal mass distribution, our hierarchical Bayesian analysis provides well-defined estimates of the key parameters characterizing the PBH population. The inferred central mass scale is $M_c = 21.44^{+0.79}_{-0.77}~M_\odot$, with a distribution width of $\sigma = 0.84^{+0.03}_{-0.03}$, a PBH abundance parameter of $\log_{10} f_{\mathrm{PBH}} = -2.67^{+0.01}_{-0.01}$, and a redshift evolution index of $\alpha = 2.19^{+0.16}_{-0.16}$. These results describe a PBH population characterized by well-localised parameter estimates, consistent with the observed BBH merger rates. For the class of log-normal mass functions considered, the inferred values of \( f_{\mathrm{PBH}} \) consistently lie below the threshold of \( 3 \times 10^{-3} \). As argued in Ref.~\cite{H_tsi_2021}, such low PBH abundances are insufficient to induce significant clustering effects, thereby justifying their exclusion from the present analysis.

\begin{figure}[h!]
    \centering
    \includegraphics[scale=0.43]{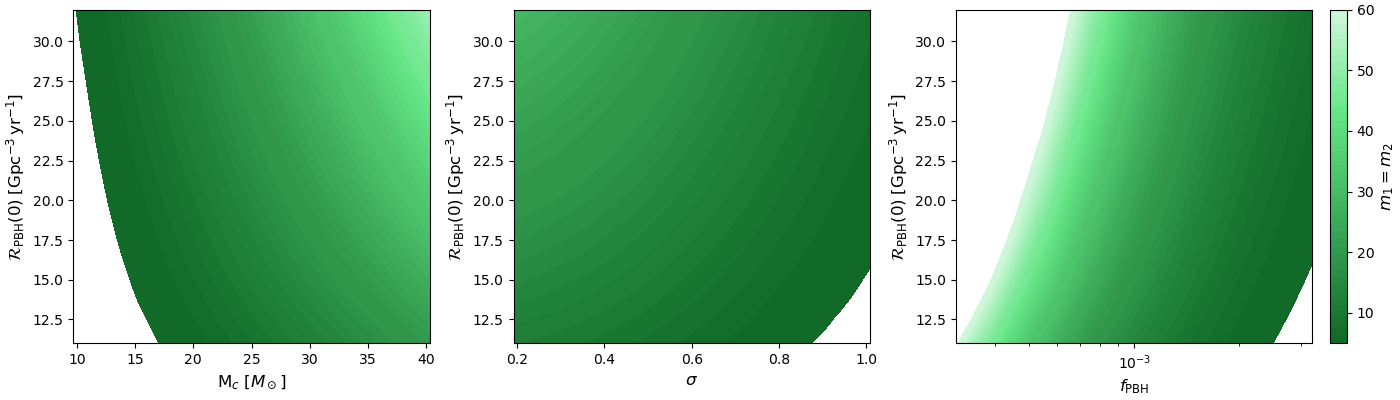}
    \caption{Dependence of the local PBH merger rate density, $\mathcal{R}_{\mathrm{PBH}}(0)$, on individual parameters of the log-normal mass function: $M_c$ (left), $\sigma$ (center), and $f_{\rm PBH}$ (right). In each panel, the remaining parameters are held fixed at representative values. The plots show the merger rate for equal-mass binaries ($m_1 = m_2$), with the color scale indicating the corresponding mass in $M_\odot$. }  \label{para_vari}
\end{figure}

Figure~\ref{para_vari} illustrates the interplay between the local PBH binary merger rate, $\mathcal{R}_{\mathrm{PBH}}(0)$, and the principal hyperparameters of the log-normal PBH mass distribution: the central mass scale $M_c$ (left panel), width $\sigma$ (middle panel), and abundance fraction $f_{\mathrm{PBH}}$ (right panel). Each panel displays the merger rate as a function of one parameter while holding the others fixed, with the color map representing the typical mass of the PBH binaries.

The left panel shows that the merger rate decreases with increasing central mass $M_c$. Lower values of $M_c$ favour higher merger rates, as smaller PBHs are both more abundant (for fixed $f_{\mathrm{PBH}}$) and more likely to form close binaries. The middle panel demonstrates that larger $\sigma$ tends to increase the merger rate, reflecting the fact that an extended distribution dilutes the population near the peak, thus lowering the frequency of optimal mass pairs for efficient binary formation. In contrast, the right panel shows a strong positive correlation between the PBH abundance $f_{\mathrm{PBH}}$ and the local merger rate, consistent with theoretical expectations: increasing the PBH fraction within the dark matter directly boosts the probability of binary formation, yielding a higher merger rate.

\begin{figure}[h!]
    \centering
    \includegraphics[scale=0.7]{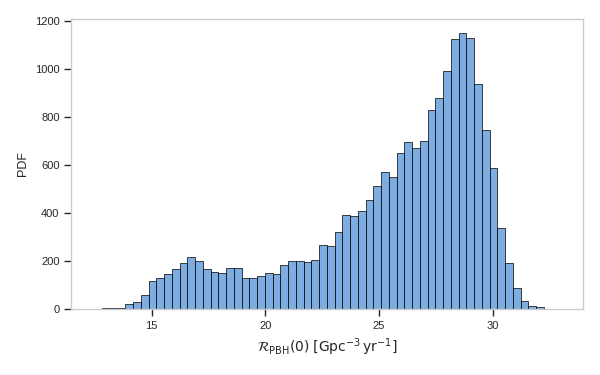}
    \caption{Posterior probability distribution of the local PBH merger rate, $\mathcal{R}_{\rm PBH}(0) [\mathrm{Gpc}^{-3}\,\mathrm{yr}^{-1}]$. }
    \label{kde}
\end{figure}

These results support the viability of the PBH scenario as a contributor to the observed GW events and motivate a detailed investigation of the associated SGWB in the following section. The predicted local PBH binary merger rate, histogram plot is shown in Figure \ref{kde},  is tightly concentrated between $23.5-30.3~ \mathrm{Gpc}^{-3}\, \mathrm{yr}^{-1}$, demonstrating excellent agreement with empirical measurements from the LVK collaborations.

The predicted SGWB spectrum shown in Figure \ref{sgwb}, $\Omega_{\mathrm{GW}}(f)$, arising from unresolved PBH binary mergers, exhibits a mean total energy density of $\langle \Omega_{\mathrm{GW}} \rangle = 1.67 \times 10^{-6}$. The maximum contribution of spectrum to the GW background is approximately at $158~\mathrm{Hz}$. SNR forecasts reveal that the ET-B would achieve SNR of $\mathcal{O}(10^{1})$ (mean SNR around $3.24$, above the considered threshold SNR $\geq$ 2) for this background, making it sensitive to the predicted PBH signal. In contrast, Advanced LIGO and Advanced Virgo are expected to reach considerably lower SNRs of $\mathcal{O}(10^{-2}) $, respectively, which are below the detection threshold for the anticipated PBH-induced SGWB.

\section{Conclusion} \label{sec6}

In this work, we have presented a comprehensive analysis of PBH binary mergers and their associated SGWB, leveraging the latest GW data from the LVK collaborations. By adopting a log-normal mass function and performing hierarchical Bayesian inference, we obtained well-localized posterior distributions for the key PBH population parameters, including the central mass scale, mass function width, abundance, and redshift evolution index. Our findings indicate that the most populated local PBH binary merger rate is within the estimated merger rate by the LVK collaboration \cite{Gwtc3}.

We inferred the central mass scale ($M_c$), the width of the mass function ($\sigma$), the PBH abundance ($\log_{10} f_{\mathrm{PBH}}$), and the redshift evolution index ($\alpha$) from the posterior distributions obtained through hierarchical Bayesian analysis. Our results reveal that the merger rate decreases with increasing central mass $M_c$, but increases strongly with the PBH abundance $f_{\mathrm{PBH}}$ and mass function width $\sigma$, with higher merger rates typically associated with lower PBH masses.  Higher merger rates are typically associated with lower PBH masses, highlighting the critical interplay between the underlying mass spectrum and the observable GW signal. The inferred PBH abundance is at the level of $\mathcal{O}(10^{-3})$, well below thresholds where additional effects become significant.

Our predictions for the SGWB, originating from unresolved PBH binary mergers, yield a mean total energy density of $\langle \Omega_{\mathrm{GW}} \rangle = 1.67 \times 10^{-6}$, with the spectrum peaking near $158~\mathrm{Hz}$. SNR forecasts indicate that, although current detectors such as Advanced LIGO and Advanced Virgo are unlikely to detect this stochastic background, the next-generation ET-B will possess the requisite sensitivity to probe the PBH contribution to the SGWB.

Overall, our results demonstrate that the PBH merger scenario not only fits current BBH merger rate observations but also produces a SGWB signal that is potentially detectable by a future GW observatory (ET-B), with the considered thresholds. This supports the ongoing search for PBHs as a component of dark matter and underscores the importance of SGWB measurements in probing the early universe.

%Our results indicate that a PBH population with an abundance \( f_{\mathrm{PBH}} \sim 10^{-3} \), characterised by a log-normal mass distribution, can account for a significant portion of the observed BBH mergers and generate a SGWB potentially within reach of current and future detectors. These findings reinforce the strong motivation for continued observational efforts to probe the PBH parameter space through both individual GW events and stochastic background searches.

\section*{Acknowledgements}
Authors also express their sincere thanks to Ranjini Mondol of for insightful discussions and comments that have enhanced the quality of the manuscript. The work of M.K. was performed in Southern Federal University with financial support from a grant of the Russian Science Foundation No 25-07-IF. 
 
\bibliographystyle{JHEP}
\bibliography{citation}
\end{document}